\documentclass[aps,twocolumn,showpacs]{revtex4}
\usepackage{graphicx}

\begin{document}

\title{Optimal dimensionality for quantum cryptography}
\author{D.B.Horoshko} \email{horoshko@dragon.bas-net.by}
\author {S.Ya.Kilin}
\affiliation{Institute of Physics, Belarus National Academy of
Sciences, F.Skarina Ave. 68, Minsk 220072 Belarus}

\begin{abstract}
We perform a comparison of two protocols for generating a cryptographic key
composed from d-valued symbols: one exploiting a string of independent
qubits and another one utilizing d-level systems prepared in states
belonging to d+1 mutually unbiased bases. We show that the protocol based on
qubits is optimal for quantum cryptography, since it provides higher
security and higher key generation rate.
\end{abstract}


\maketitle

Quantum cryptography is a technology allowing two distant parties
to share a random string of symbols (cryptographic key) in such a
way that any eavesdropping from the third party will cause a
detectable disturbance. This can be achieved by using for
communication quantum systems in states highly sensitive to
measurement. Since its discovery by Bennett and Brassard (BB84
protocol) \cite{BB84}, quantum cryptography plays a central role
in the field of quantum information processing \cite{QI}, being
the most experimentally advanced application in this field and
serving as a 'testing ground' for new ideas \cite{Gisin-Review}.
In particular, recently it has been shown that the exploiting of
quantum systems of dimension higher than two (qudits) results in
higher disturbance cost of information, making the protocol
tougher to eavesdrop \cite{BP-T,BP-P,Bour,Cerf,BM}. This situation
differs dramatically from that of classical information, where the
length of the alphabet does not play any role, and therefore the
most simple binary encoding is commonly used \cite{Shannon}.

However, the comparison between qubit and qudit protocols seems to
us to be incomplete, since the qudit encoding is not compared to
the case where a dit (d-valued symbol) is encoded in a sequence of
independent qubits. Indeed, to share a key composed from dits (for
simplicity we consider $d=2^n$) the communicating parties can use
the standard binary cryptographic protocol with the corresponding
mapping of classical data, e.g. a four-valued symbol (quart) can
be decomposed into two bits, an eight-valued symbol - in three
bits et cetera.

In the present paper our aim is to consider the protocol for distributing a $%
2^n$-letter cryptographic key by means of 6-state protocol for qubits \cite
{Bruss,BP-G} and to compare it to the protocol utilizing $2^n$-level quantum
systems with states chosen from mutually unbiased bases \cite
{BP-T,BP-P,Bour,Cerf,BM}. Our comparison will be significant for two
practical problems. On one hand, if one has two-level systems, such as
single photons with different polarizations, one can compose a $2^n$-level
system from a sequence of $n$ qubits, but the access to any state in the
Hilbert space of such a system will require entanglement between the qubits,
i.e. an additional quantum resource. Our analysis will show, how much one
gains with respect to security and the key generation rate in exchange to
this resource. On the other hand, if one has a $2^n$-level system, one can
represent its Hilbert space as a direct product of $n$ two-dimensional
systems, and use it for standard cryptography based on qubits, restricting
oneself to factorizing states of $n$ subsystems. Our analysis should show if
this strategy yields any benefice.

Let us first recall the details and notations of the 6-state extension of
BB84 protocol for quantum cryptography \cite{Bruss,BP-G}. In this protocol
the sender (Alice) generates a random bit and encodes it in a state of a
qubit, choosing randomly one of three possible bases. The first basis is
denoted as $\left| 0\right\rangle $ and $\left| 1\right\rangle $. The other
two are
\begin{equation}
\left| 0^{\prime }\right\rangle =\frac 1{\sqrt{2}}\left( \left|
0\right\rangle +\left| 1\right\rangle \right) ,\qquad \left| 1^{\prime
}\right\rangle =\frac 1{\sqrt{2}}\left( \left| 0\right\rangle -\left|
1\right\rangle \right) ,  \label{1}
\end{equation}
and
\begin{equation}
\left| 0^{\prime \prime }\right\rangle =\frac 1{\sqrt{2}}\left( \left|
0\right\rangle +i\left| 1\right\rangle \right) ,\qquad \left| 1^{\prime
\prime }\right\rangle =\frac 1{\sqrt{2}}\left( \left| 0\right\rangle
-i\left| 1\right\rangle \right) .  \label{2}
\end{equation}
The qubit is sent to the receiver (Bob), who measures it in a randomly
chosen basis. This procedure is repeated many times. After the public
disclosure of bases, chosen by Alice and Bob, both parties keep only those
bits for which the bases coincide. In this way Alice and Bob generate a
shared random string of bits. A part of this string can be communicated via
public channel to determine Bob's disturbance - the percentage of
incorrectly received bits. On the basis of the measured disturbance a
standard procedure of privacy amplification is performed in order to
decrease the information of possible eavesdropper \cite{Privacy}. As a
result, a secure shared cryptographic key is generated.

The same protocol can be used for generating a cryptographic key composed
from $2^n$-valued symbols. Alice generates a random string of symbols,
writes each symbol in binary and communicates with Bob as described above.
After the binary key is generated, each $n$ successive bits are mapped back
into $2^n$-valued symbols. Now we need to consider how this protocol is
affected by intervention of an eavesdropper (Eve). We restrict ourselves to
individual attacks, for which at most one qubit at a time is attacked.

Eve can exploit different strategies for individual eavesdropping,
e.g., the simple intercept-resend strategy, where she intercepts a
qubit, measures it in a random basis and resends to Bob a qubit,
prepared in the measured state. In this way Eve gains some
information in exchange to disturbance introduced into Bob's key.
In order not to be detected, Eve is interested in increasing her
information and decreasing the Bob's disturbance. The best known
at present strategy for that end is the asymmetric cloning of
qubit \cite{Bour,Cerf,BM}. In this type of attack Eve attaches to
the Alice's qubit two additional qubits denoted $E$ and $M$,
performers a unitary transformation of all three qubits, and sends
the first qubit further to Bob, keeping the other two. After the
disclosure of bases used by Alice and Bob, Eve measures her two
qubits in the basis chosen by Alice. The cloning transformation,
written in ''correct'' basis, is \cite{Bour}:
\begin{eqnarray}
\left| \psi _k\right\rangle _A &\longrightarrow &\left| \psi _k\right\rangle
_B\left( \frac \alpha {\sqrt{2}}\left( \left| 0\right\rangle _E\left|
0\right\rangle _M+\left| 1\right\rangle _E\left| 1\right\rangle _M\right)
\right.  \label{3} \\
&&\left. +\frac \beta {\sqrt{2}}\left| \psi _k\right\rangle _E\left| \psi
_k\right\rangle _M\right)  \nonumber \\
&&+\frac \beta {\sqrt{2}}\left| \psi _{k+1}\right\rangle _B\left| \psi
_k\right\rangle _E\left| \psi _{k+1}\right\rangle _M  \nonumber
\end{eqnarray}
where the two states of the basis are denoted as $\left| \psi
_0\right\rangle =\left| 0\right\rangle $ and $\left| \psi _1\right\rangle
=\left| 1\right\rangle $, and the summation in index is taken modulo 2. The
real positive numbers $\alpha $ and $\beta $ are parameters of the cloning
machine, they satisfy the relation
\begin{equation}
\alpha ^2+\alpha \beta +\beta ^2=1.  \label{4}
\end{equation}
The first term in the r.h.s. of Eq. (\ref{3}) corresponds to Bob having no
error, while the second term corresponds to a flip of Bob's qubit. After the
measurement of her qubits Eve accepts the value of the measurement of qubit $%
E$ as the corresponding bit in the eavesdropped key. There are
three possible outcomes of measurements by Eve and Bob of the
state described by Eq. (\ref{3}): (i) Eve and Bob receive correct
values of Alice's bit with probability $p_0=\left( \alpha +\beta
\right) ^2/2$; (ii) Bob receives the
correct value, but Eve receives the incorrect one with probability $%
p_e=\alpha ^2/2$; and, finally, (iii) Bob receives the incorrect value,
while Eve receives the correct one with probability $p_b=\beta ^2/2$. Bob in
no way can guess which of the cases takes place (without revealing his bit
to Alice). In the meanwhile, Eve can distinguish first two cases from the
third one by comparing the outcomes of measurements of her two qubits $E$
and $M$. Let us denote the difference (modulo 2) of these outcomes as $m$.
If the outcomes coincide ($m=0$), then the first or the second case occurs,
if they differ ($m=1$), then the third one takes place. Given $m=0$ Eve's
probability to get the correct value is $q=p_0/(p_0+p_e)$. On the basis of
these probabilities one can calculate information got by Bob and Eve. We
perform this calculation for keys composed from $2^n$-valued symbols.

First we consider the case of $n=2$. In this case two qubits are used for
sending a 4-valued symbol (quart). Eve clones each qubit separately,
measures her qubits in correct basis and keeps the values of $E$ qubits for
her key. Besides she calculates differences of outcomes of measuring her
qubits $E$ and $M$ - $m_1$ for the first qubit and $m_2$ for the second one.
Four possible cases should be distinguished depending on the values of $m_1$
and $m_2$. Let us suppose, without loss of generality, that the value 00 was
sent by Alice. Then the four cases are as follows.

For $m_1=m_2=0$, which happens with probability $\xi _{00}=(p_0+p_e)^2$, the
distribution of Eve's quart is $P_E^{(00)}=(q^2,q(1-q),q(1-q),(1-q)^2)$,
while that of Bob's one is $P_B^{(00)}=(1,0,0,0)$. Here we write the
distribution in the vector form $(p_{00},p_{01},p_{10},p_{11})$, where $%
p_{ij}$ denotes the probability for quart to have value $ij$ in binary
notation.

Similarly, for $m_1=0$, $m_2=1$ (first bit being the major one), which
happens with probability $\xi _{01}=p_b(p_0+p_e)$, we have $%
P_E^{(01)}=(q,0,1-q,0)$, $P_B^{(01)}=(0,1,0,0)$.

For $m_1=1$, $m_2=0$, which happens with probability $\xi _{10}=p_b(p_0+p_e)$%
, we have $P_E^{(10)}=(q,1-q,0,0)$, $P_B^{(10)}=(0,0,1,0)$.

And, finally, for $m_1=m_2=1$, which happens with probability $\xi
_{11}=p_b^2$, we have $P_E^{(10)}=(1,0,0,0)$, $P_B^{(10)}=(0,0,0,1)$.

Now we are ready to calculate the information on Alice's quart received by
Bob and Eve. This information is calculated as $I=2-H(P)$, where $H(P)$ is
the Shannon entropy function calculated for the distribution $P$:
\begin{equation}
H(P)=-\sum_{i,j=0,1}p_{ij}\log _2p_{ij}.  \label{8}
\end{equation}
However, averaging over $m_1$ and $m_2$ is performed differently for Bob and
Eve. Bob does not know the values of $m_1$ and $m_2$, and therefore for him
the average distribution $\left\langle P_B\right\rangle =\sum \xi
_{ij}P_B^{(ij)}$ is found first, and then the information for this
distribution is calculated $I_B=2-H(\left\langle P_B\right\rangle )$. Eve,
on the contrary, knows the values of $m_1$ and $m_2$ and her information is
calculated for each case and then averaged: $I_E=2-\left\langle
H(P_E)\right\rangle $. In this way we obtain
\begin{eqnarray}
\left\langle P_B\right\rangle &=&\left(
(1-p_b)^2,p_b(1-p_b),p_b(1-p_b),p_b^2\right) ,  \label{9} \\
I_B &=&2-2h(p_b),  \label{10} \\
I_E &=&2-2h(q)(1-p_b),  \label{11}
\end{eqnarray}
where $h(q)\equiv -q\log _2q-(1-q)\log _2(1-q)$. Informations of Bob and Eve
are exactly twice that for the case of bits, as could be expected. If one
considers the dependence of $I_B$ and $I_E$ from $p_b$ (note that $q$ can be
expressed through $p_e$, using Eq. (\ref{4})), one finds that $I_B$
decreases with $p_b$ while $I_E$ increases. The intersection of two curves
defines the border value of $p_b$, below which Bob's information exceeds
Eve's one and therefore the standard procedure of privacy amplification if
applicable. This value is the solution of equation
\begin{equation}
2-2h(p_b)=2-2h(q)(1-p_b),  \label{12}
\end{equation}
and is the same as for the 6-state protocol for quantum cryptography with
qubits, $p_b=0.1564$. In the case of generating binary key $p_b$ has the
sense of disturbance, i.e. probability for Bob to get incorrect symbol in
his key. In the case of cryptography with quarts, considered here, the
intersection of two curves occurs at the same value of $p_b$, but the border
disturbance is now $\tilde{D}^{(4)}=1-(1-p_b)^2=0.2883$. The growth of
disturbance reflects the fact that an incorrect quart can be received when
only one bit is wrong and another one is correct. This point was mentioned
in Ref. \cite{BP-G}, but did not deserve a detailed consideration up to now.

The considered cryptographic protocol for generating a key
consisting from quarts can be compared to the protocol where 5
mutually unbiased bases in four-dimensional Hilbert space are used
\cite{Bour,Cerf,BM}. In the latter protocol Alice and Bob use for
encoding such bases that the overlap of two any states from
different bases is the same. The rest of the protocol resembles
BB84. The use of mutually unbiased bases guarantees that no
information on the encoded symbol is got if an incorrect basis is
chosen for measurement. Therefore the protocol implementing such
bases should provide the best security for cryptography with
qudits. The analysis of eavesdropping by means of asymmetric
cloning of qudits shows that in this case \cite{Bour,Cerf}
\begin{eqnarray}
\left\langle P_B\right\rangle  &=&\left( 1-D,D/3,D/3,D/3\right) ,  \label{13}
\\
I_B &=&2+(1-D)\log _2(1-D)+D\log _2\frac D3,  \label{14} \\
I_E &=&2+(1-D)(1-\mu )\log _2(1-\mu )  \label{15} \\
&&+(1-D)\mu \log _2\frac \mu 3,  \nonumber
\end{eqnarray}
where $\mu =D_E/(1-D)$, $D$ and $D_E$ being Bob's and Eve's disturbances
respectively, connected by the following parametric relations: $D_E=3\bar{%
\alpha}^2/4$, $D=3\bar{\beta}^2/4$, where $\bar{\alpha}$ and $\bar{\beta}$
are real positive parameters of the cloning machine for four-level quantum
systems, satisfying the normalization relation
\begin{equation}
\bar{\alpha}^2+\frac 12\bar{\alpha}\bar{\beta}+\bar{\beta}^2=1.  \label{18}
\end{equation}
The solution of equation $I_B=I_E$ in this case gives the result $%
D^{(4)}=0.2666$. We see that $D^{(4)}<\tilde{D}^{(4)}$, that is the
cryptographic protocol exploiting qubit pairs for generating a key of quarts
is more secure against eavesdropping attacks, than the protocol, utilizing
four-level systems prepared in one of 5 mutually unbiased bases. Besides,
the former protocol provides higher key generation rate, since $1/3$ of
systems is used for the key, against $1/5$ in the latter case.

This result is easily generalized to any value of $n$. If a sequence of $n$
qubits is used for generating a key of $2^n$-valued symbols, then Bob's
disturbance corresponding to intersection of informational curves for Bob
and Eve is $\tilde{D}^{(2^n)}=1-(1-0.1564)^n$. The corresponding disturbance
$D^{(2^n)}$ for the case where $2^n$-dimensional systems are used, being
prepared in a state chosen from $2^n+1$ mutually unbiased bases, is
calculated as a solution of equation $I_B=I_E$ where \cite{Bour,Cerf}
\begin{eqnarray}
I_B &=&n+(1-D)\log _2(1-D)+D\log _2\frac D{2^n-1},  \label{19} \\
I_E &=&n+(1-D)(1-\mu )\log _2(1-\mu )  \label{20} \\
&&+(1-D)\mu \log _2\frac \mu {(2^n-1)},  \nonumber
\end{eqnarray}
where again $\mu =D_E/(1-D)$, but this time $D$ and $D_E$ are connected by
equations $D_E=(1-2^{-n})\tilde{\alpha}^2$, $D=(1-2^{-n})\tilde{\beta}^2$,
where
\begin{equation}
\tilde{\alpha}^2+\frac 1{2^{n-1}}\tilde{\alpha}\tilde{\beta}+\tilde{\beta}%
^2=1.  \label{23}
\end{equation}
The both disturbances $\tilde{D}^{(2^n)}$ and $D^{(2^n)}$, calculated
numerically, are plotted in Fig.1 as functions of $n$. The plot shows, that $%
\tilde{D}^{(2^n)}$ is always greater than $D^{(2^n)}$, i.e. a protocol based
on qubits is more secure than that based on qudits. Note that for symbols
with more than two values the disturbance is not limited by 1/2.

\begin{figure}[h]
\centering
\includegraphics[width=0.45\textwidth]{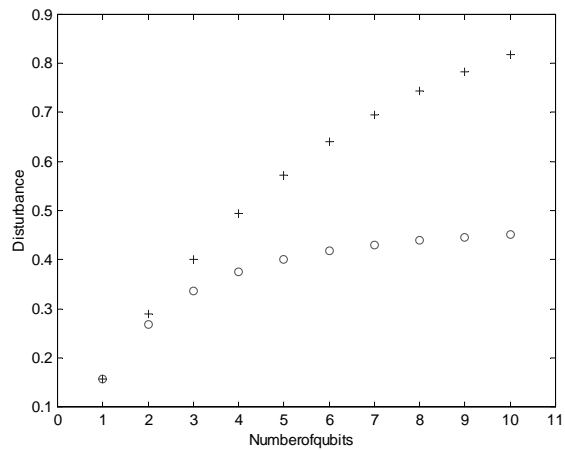}
\caption{The border disturbance for sharing a key composed from
$2^n$-valued symbols, using qubits (crosses) and $2^n$-level
systems (circles).}
\end{figure}

One could conjecture that a sequence of qubits encoding one quart could be
effectively eavesdropped by a collective attack. But this is not so, since
any collective attack produces correlation between attacked qubits, and
therefore can be easily detected by Bob. Undetectable collective attack
should address randomly chosen qubits, i.e. it may to be expected to be
equally successful for qubits and qudits. Moreover, it is still unclear at
present, if collective attacks are more effective than individual ones \cite
{Gisin-Review}.

We conclude that the optimal dimensionality for quantum
cryptography is 2, i.e. two-dimensional systems are the best tool
for information encoding, like in the classical information
theory. This result leads us to two practical recommendations
concerning implementation of quantum cryptography. First, if one
has two-level systems available for quantum cryptography, there is
no sense in entangling them for producing systems of higher
dimensions. Second, if one has multilevel systems at hand, it is
more profitable to decompose the states of each system into direct
product of two-dimensional spaces and use them for standard
cryptographic protocol with qubits. This strategy provides better
security and better key generation rate.

Authors gratefully acknowledge support from INTAS, Open call 2001
project 2097 and from Belorussian Republican Foundation for
Fundamental Research.

\end{document}